\documentclass[twocolumn,showpacs,preprintnumbers,amsmath,amssymb]{revtex4}
\usepackage{graphicx}% Include figure files
\usepackage{dcolumn}% Align table columns on decimal point
\usepackage{bm}% bold math

\begin{document}

\preprint{}

\title{Conditions on local operations to create quantum correlation}% Force line breaks with \\

\author{Xueyuan Hu}
\email{xyhu@iphy.ac.cn}
% \altaffiliation[Also at ]{Physics Department, XYZ University.}%Lines break automatically or can be forced with \\
\author{Heng Fan}
\author{Duan-Lu Zhou}
\author{Wu-Ming Liu}
\affiliation{Beijing National Laboratory for Condensed Matter
Physics, Institute of Physics, Chinese Academy of Sciences, Beijing
100190, China}
\date{\today}% It is always \today, today,
             %  but any date may be explicitly specified

\begin{abstract}
We prove the condition for local trace-preserving channels to create
quantum correlation from initially classical states. For two-qubit
states, the necessary and sufficient condition for a channel that
cannot create quantum correlation in any initially classical state
is that it is either a completely decohering channel or a mixing
channel. For states in higher dimensions, we show that even a mixing
channel can create quantum correlation from a classically correlated
state. Our results reveal that mixedness is important for creation
of quantum correlation in high-dimension quantum systems.
\end{abstract}

% PACS, the Physics and Astronomy
                             % Classification Scheme.
%\keywords{Suggested keywords}%Use showkeys class option if keyword
                              %display desired
\maketitle

% Put \label in argument of \section for cross-referencing
%\section{\label{}}
Quantum correlation is the unique phenomenon of quantum physics and
believed to be a resource for quantum information processes which
can generally surpass the corresponding classical schemes. Much of
the previous studies focus on entanglement, a well-known quantum
correlation, since its apparent role in teleportation, superdense
coding \cite{tele,PhysRevA.54.1869}, etc. Recently, quantum discord,
a different measure of the nonclassicalness of correlation, began to
attract much attention since the discovery that some quantum
information schemes can be realized without entanglement but with a
positive quantum discord
\cite{PhysRevLett.100.050502,PhysRevLett.107.080401}. And much
progress has been made in studying quantum discord, to quantify the
amount of quantum discord in some physical systems and to give it
some intuitive and operational interpretations. For example, it is
shown that quantum discord can be operational interpreted as the
minimum information missing from the environment
\cite{arXiv:1110.1664v1}.

Some other measures of quantum correlation are also of broad
interest. Quantum deficit \cite{PhysRevLett.89.180402} is
operationally defined as the following. Work can be extracted from a
quantum system using Szilard's engine. When composite systems are
considered, the extracted work using global operations is different
from that when only local operation and classical communications
(LOCC) are allowed. The minimum difference is called quantum
deficit. One-way quantum deficit is defined as the quantum deficit
when only one-way communication is allowed \cite{RevModPhys.81.1}.
One-way quantum deficit has been found as the reason for
entanglement irreversibility \cite{PhysRevLett.107.020502} and can
be related to quantum entanglement via an interesting scheme
\cite{PhysRevLett.106.160401,PhysRevLett.106.220403}.

Quantum noise usually plays a destructive role in quantum
information process. However, there are situations that local
quantum noise can enhance nonlocal quantum properties for some mixed
quantum states. For example, local amplitude damping can increase
the average teleportation fidelity for a class of entangled states
\cite{PhysRevA.62.012311,PhysRevA.78.022334,PhysRevA.81.054302}.
Quantum discord can also be increased or created by local noise
\cite{PhysRevA.84.022113}. An interesting result is that any
separable state with positive quantum discord can be produced by
local positive operator-valued measure (POVM) on a classical state
in a larger Hilbert space \cite{PhysRevA.78.024303}. In fact, almost
all states in the Hilbert space contains quantum discord, and an
arbitrary small disturbance can drive a classical state into a
quantum state with nonzero quantum deficit
\cite{PhysRevA.81.052318}. Counterintuitively, it has recently been
discovered that mixedness is as important as entanglement for
quantum correlation. In particular, some mixed states contain more
quantum discord than that of maximally entangled pure state when the
dimension of the system is large enough
\cite{PhysRevLett.106.220403}. Thus it is of interest to know how is
the effect of mixedness on the quantum correlation of quantum
states.

In this article, we will concentrate on the question: what kind of
local noise can create quantum correlation in initially zero-discord
states? Apparently, completely decohering channels, which takes any
input state into a diagonal state, cannot create quantum discord in
any input state. Then we define a class of channels as mixing
channels which does not suppress the entropy of any quantum state.
We prove that for two-qubit states, a channel that cannot create
quantum deficit from any classical state if and only if it is either
a mixing channel or a completely decohering channel. For states in
higher dimension, even mixing channel can create quantum discord.
This result provides a strong evidence that mixedness is important
for the quantum correlation in high-dimension quantum states.

A quantum channel can be described by a trace-preserving completely
positive map
\begin{equation}
\rho\rightarrow\Lambda(\rho)=\sum_i\mathrm E^{(i)}\rho\mathrm
E^{(i)\dagger},
\end{equation}
where $\mathrm E^{(i)}$'s are called Kraus operators satisfying
$\sum_i\mathrm E^{(i)\dagger}\mathrm E^{(i)}=\mathrm I$. The above
expression can be obtained by employing an ancillary system $R$
doing a global unitary operation and then tracing out $R$
\begin{equation}
\Lambda(\rho)=\mathrm{Tr}_R[U(\rho\otimes|0\rangle_R\langle0|)U^{\dagger}],\label{Kraus}
\end{equation}
and $\mathrm E^{(i)}=_R\langle i|U|0\rangle_R$. Notice that a
unitary operator $U'$ on the ancillary system $R$ does not change
the effect of the channel, so $\mathrm
E^{\prime(i)}=\sum_jU'_{kj}E^{(j)}$ describes the same channel as in
Eq. (\ref{Kraus}) for any unitary operator $U'$.

In the following, we will define a class of quantum channels as the
mixing channels. A mixing channel $\Lambda^{\mathrm M}$ acting on a
system $S$ never decreases the entropy of $S$ for arbitrary input
state $\rho_S$
\begin{equation}
S(\Lambda^{\mathrm M}(\rho_S))\geq S(\rho_S),\ \forall
\rho_S.\label{mixing_channel_d}
\end{equation}
It is worth mentioning that when a channel is a mixing channel, its
extension to larger systems $\mathrm I_A\otimes\Lambda_S^{\mathrm
M}$ is still a mixing channel. Consider an initial state
$\rho_{AS}^0$, we will prove that
\begin{equation}
S(\mathrm I\otimes\Lambda^{\mathrm M}_S(\rho_{AS}^0))\geq
S(\rho_{AS}^0).\label{mixing_channel_e}
\end{equation}
Employing an ancillary system $R$ such that the channel can be
written as $\mathrm I\otimes\Lambda^{\mathrm
M}_S(\rho_{AS}^0)=\mathrm{Tr}_R(\rho_{ASR})$, where
$\rho_{ASR}=\mathrm I_A\otimes
U_{SR}\rho_{AS}^0\otimes|0\rangle_R\langle0|\mathrm I_A\otimes
U_{SR}^{\dagger}$. From the strong subadditivity of entropy, we have
$S(\rho_{AS})-S(\rho_{ASR})\geq S(\rho_R)-S(\rho_{SR})\geq0$, where
the second inequality is from Eq. (\ref{mixing_channel_d}). Since
$S(\rho_{ASR})=S(\rho_{AS}^0)$, we finally arrive at Eq.
(\ref{mixing_channel_e}). We will show that in two-qubit case,
$\Lambda^{\mathrm{M}}$ is the only class of channel that cannot
increase one-way quantum deficit for any state, while in general
cases, some channels in the form of $\Lambda^{\mathrm{M}}$ can
create quantum deficit for some state. Before we proceed, let us
look more closely the explicit form of a mixing channel.

Lemma 1: a channel acting on a qubit is a mixing channel if and only
if its Kraus operator satisfies
\begin{equation}
\sum_i\mathrm E^{(i)}\mathrm E^{(i)\dagger}=\mathrm
I.\label{mixing_channel}
\end{equation}

Proof: when Eq. (\ref{mixing_channel}) is not satisfied, we consider
$\rho^{\mathrm I}=\mathrm I/2$ as the input state, and then
$\rho'=\sum_i\mathrm E^{(i)}(\mathrm I/2)\mathrm
E^{(i)\dagger}\neq\mathrm I/2$. Notice that $\rho^{\mathrm I}$ is
the maximally mixing state of a qubit, so $S(\rho')\leq
S(\rho^{\mathrm I})$. It means that the entropy is suppressed.

On the other hand, when Eq. (\ref{mixing_channel}) is satisfied, we
consider an arbitrary input state
$\rho=\sum_{i=0}^1p_i|\psi_i\rangle\langle\psi_i|$, where
$|\psi_i\rangle=u|i\rangle$ for $i=0,1$. We have
\begin{eqnarray}
&&[\sum_{i}\mathrm E^{(i)}|\psi_0\rangle\langle \psi_0|\mathrm
E^{(i)\dagger},\sum_{i}\mathrm
E^{(i)}|\psi_1\rangle\langle\psi_1|\mathrm
E^{(i)\dagger}]\nonumber\\
&=&-[\sum_i\mathrm E^{(i)}\mathrm E^{(i)\dagger},\sum_i\mathrm
E^{(i)}u\sigma^zu\mathrm E^{(i)\dagger}]=0.
\end{eqnarray}
Therefore, $\sum_i\mathrm E^{(i)}|\psi_0\rangle\langle
\psi_0|\mathrm E^{(i)\dagger}$ and $\sum_i\mathrm
E^{(i)}|\psi_1\rangle\langle \psi_1|\mathrm E^{(i)\dagger}$ share
common eigenstates $\{|\phi_j\rangle\}$. Let $\lambda_j^k$ ($k=0,1$)
be the corresponding eigenvalues, and then from the trace-preserving
property
\begin{equation}
\lambda_0^k+\lambda_1^k=1.\label{l1}
\end{equation}
Meanwhile, from Eq. (\ref{mixing_channel}) we have
$\sum_{j=0}^1(\sum_{k=0}^1\lambda_j^k)|\phi_j\rangle\langle\phi_j|=\mathrm
I$ and then
\begin{equation}
\lambda_j^0+\lambda_j^1=1.\label{l2}
\end{equation}
Combining Eqs. (\ref{l1}) and (\ref{l2}), we have
$\lambda_0^0=\lambda_1^1\equiv(1+\lambda)/2$ and
$\lambda_0^1=\lambda_1^0\equiv(1-\lambda)/2$ ($|\lambda|\leq1$).
Therefore,
\begin{eqnarray}
S(\rho')&=&S(\sum_{k=0}^1\rho_k\sum_i\mathrm
E^{(i)}|\psi_k\rangle\langle\psi_k|\mathrm E^{(i)\dagger})\nonumber\\
&=& H(\frac{1+p\lambda}{2})\geq H(\frac{1+p}{2})=S(\rho),
\end{eqnarray}
where $p_0\equiv(1+p)/2$. It means that the channel satisfying Eq.
(\ref{mixing_channel}) never suppresses the entropy for any input
state, completing the proof of Lama 1. This necessary and sufficient
condition for mixing channel was also proved in Ref.
\cite{PhysRevA.82.052103} using a different method.

Quantum correlation can be measured by one-way quantum deficit. For
a bipartite quantum state $\rho_{AB}$, when global operations are
allowed, $1-S(\rho_{AB})$ amount of work can be extracted. When the
Maxwell's demon is at site $A$ and one-way communication from $B$ to
$A$ is allowed, the demon can recover the state as
\begin{equation}
\rho_{AB}^{\mathrm
D}=\sum_i\tilde{\rho}_A^i\otimes\Pi_B^i,\label{half_classical}
\end{equation}
where $\{\Pi_B^i\}$ is basis of von Neumann measurement on $B$, and
$\tilde{\rho}_A^i$ is the unnormalized density matrix of $A$ when
measurement result $i$ is obtained. Therefore, only
$1-S(\rho_{AB}^{\mathrm D})$ amount of work can be extracted in this
situation. The minimum difference of extracted work is one-way
quantum deficit \cite{PhysRevA.67.012320,RevModPhys.81.1}
\begin{equation}
\Delta^{\leftarrow}_{AB}\equiv\min_{\Pi_B^i}S(\rho_{AB}^{\mathrm
D})-S(\rho_{AB}).
\end{equation}
Ref. \cite{PhysRevA.67.012320} discussed that one-way quantum
deficit is the thermaldynamical interpretation for quantum discord.
However, it differs from the usually mentioned concept ``quantum
discord'', which is defined as \cite{PhysRevLett.88.017901}
\begin{equation}
\delta^{\leftarrow}_{AB}\equiv\min_{\Pi_B^i}S_{A|B}(\rho_{AB}^{\mathrm
D})-S_{A|B}(\rho_{AB}).
\end{equation}
Generally, $\delta^{\leftarrow}_{AB}\leq\Delta^{\leftarrow}_{AB}$.
However, both quantum deficit and quantum discord vanish if and only
if the state is a half-classical state in the form of Eq.
(\ref{half_classical}). We here derive a simple necessary and
sufficient condition for vanishing quantum discord in a separable
state $\xi_{AB}=\sum_ip_i\xi_i^A\otimes\xi_i^B$:
\begin{equation}
[\xi_i^B,\xi_j^B]=0,\ \forall i,j,\label{criterion1}
\end{equation}
which is in accordance with the necessary and sufficient condition
proposed in Ref. \cite{PhysRevLett.105.190502} for separable states.
Eq. (\ref{criterion1}) is equivalent to that $\xi_i$ and $\xi_j$
share common eigenvectors. By choosing these eigenvectors as the
basis for von Neumann measurement, the state does not change after
the measurement, which is in turn equivalent to that $\xi_{AB}$ is a
half-classical state.

Now we are ready to prove the following theorem, which is the main
result of this paper.

Theorem 1: a local channel $\Lambda$ acting on qubit $B$ of the
two-qubit half-classical state
$\rho_{AB}=\sum_{ij}\tilde{\rho}^A_j\otimes\Pi_j$ never creates
quantum deficit if and only if $\Lambda$ is one of the following two
cases:

Case 1: $\Lambda$ is a mixing channel;

Case 2: $\Lambda$ is a completely decohering channel. It means that
$\Lambda$ can be decomposed as $\mathrm E^{(i)}=u\mathrm
E^{\prime(i)}u'$, where $u$ and $u'$ are some unitary operators, and
$\Lambda'(\cdot)\equiv\sum_iE^{\prime(i)}(\cdot)E^{\prime(i)\dagger}$
takes any input state to a diagonal state on basis
$\{|0\rangle,|1\rangle\}$.

Proof: by using Eq. (\ref{criterion1}), it is equivalent to prove
that
\begin{equation}
[\sum_{i}\mathrm E^{(i)}|\phi\rangle\langle\phi|\mathrm
E^{(i)\dagger},\sum_{i}\mathrm
E^{(i)}|\psi\rangle\langle\psi|\mathrm E^{(i)\dagger}]=0\label{c3}
\end{equation}
holds for any orthogonal states $|\phi\rangle$ and $|\psi\rangle$ of
qubit B if and only if $\mathrm E^{(i)}$ satisfy Eq.
(\ref{mixing_channel}). Substituting $|\phi\rangle=u|0\rangle$ and
$|\psi\rangle=u|1\rangle$ in Eq. (\ref{c3}) gives
\begin{equation}
[\sum_i\mathrm E^{(i)}\mathrm E^{(i)\dagger},\sum_j\mathrm
E^{(j)}u\sigma^zu^{\dagger}\mathrm E^{(j)\dagger}]=0,\ \forall
u,\label{c4}
\end{equation}
which is equivalent to
\begin{equation}
[\sum_i\mathrm E^{(i)}\mathrm E^{(i)\dagger},\sum_j\mathrm
E^{(j)}\rho\mathrm E^{(j)\dagger}]=0,\ \forall \rho.\label{c5}
\end{equation}
Eq. (\ref{c5}) holds only for the following two cases:

Case 1: $\sum_i\mathrm E^{(i)}\mathrm E^{(i)\dagger}=\mathrm I$. It
means that the channel is a mixing channel.

Case 2: the two matrices $\sum_i\mathrm E^{(i)}\mathrm
E^{(i)\dagger}$ and $\sum_j\mathrm E^{(j)}\rho\mathrm
E^{(j)\dagger}$ can be diagonized by the same unitary operation,
which we label as $U$. Setting $\mathrm E^{\prime(i)}=U\mathrm
E^{(i)}$, we can imply that the channel $\Lambda'$ takes any input
state to a diagonal form. Completing the proof.

We call channels in Case 2 the completely decohering channels since
they takes any input state to a completely decohered state. Some of
the complete decoherence channels, such as dephasing or depolarizing
in $t\rightarrow\infty$ limit, are mixing channels. However, there
are situations that a complete decoherence channel is not a mixing
channel. For example, completely amplitude damping takes any state
to a pure ground state with zero entropy. Although amplitude damping
can create quantum discord in finite time, the created quantum
discord vanishes asymptotically for $t\rightarrow\infty$. Notice
that Case 2 takes any state to a half-classical state. Therefore, we
can also call it a quantum correlation breaking channel.

While Case 2 corresponds to the $t\rightarrow\infty$ limit of a
dynamics of an open system, Case 1 can describe the whole process of
some dynamics. When a dynamics increases the entropy of system
monotonically for any input state, it can not create quantum discord
for any two-qubit input state during the whole evolution; otherwise,
the dynamic can create some quantum discord for some input state.

Now let us look more closely about the dynamics that cannot create
quantum discord for any two-qubit input state. From Theorem 1, such
dynamics increases the entropy of qubit monotonically for any input
state; otherwise, the quantum discord can be created when the
entropy is suppressed. Notice that non-Markovian dynamics involves
the information about the system in an earlier time flowing from
environment to system, which may cause suppression of entropy, so
any non-Markovian dynamics can create quantum discord for some input
half-classical states at finite times. Furthermore, not all
Markovian process can keep the classicalness of any two-qubit state.
For example, the Markovian amplitude damping, with Kraus operators
\begin{equation}
\mathrm E^{(0)}=\left(\begin{array}{cc}1 & 0\\
0 & \sqrt{1-p}\end{array}\right), \mathrm E^{(1)}=\left(\begin{array}{cc}0 & \sqrt{p}\\
0&0\end{array}\right)
\end{equation}
can create quantum discord for input state
$\rho=\tilde{\rho}_1^A\otimes|+\rangle\langle+|+\tilde{\rho}_2^A\otimes|-\rangle\langle-|$.
This can be explained as the information of the environment itself
flowing to the system, causing the suppression of entropy and
consequently, the creation of quantum discord. For general Markovian
processes, described as the following master equation
$d\rho/dt=\mathrm L\rho$, where
\begin{equation}
\mathrm
L\rho=-i[H,\rho]+\sum_{\alpha,\beta}\gamma_{\alpha\beta}[F_{\alpha}\rho
F_{\beta}^{\dagger}-\frac12\{F_{\beta}^{\dagger}F_{\alpha},\rho\}_+],\label{lindblad_operator}
\end{equation}
and $\{F_{\alpha}\}$ constitutes a complete basis for the vector
space of bounded operators acting on the Hilbert space of qubit $B$.
Without loss of generality, $\{F_{\alpha}\}$ can be chosen as
$\{\mathrm I,\sigma_x,\sigma_y,\sigma_z\}$. Since $\mathrm L$ does
not depend on time $t$, we can write $\rho(t+dt)=\Lambda(\rho(t))$
with $\Lambda$ independent of $t$. Therefore, the necessary and
sufficient condition for $\mathrm L$ to preserve the classicality of
any two-qubit half-classical state is that
\begin{equation}
\mathrm L(\mathrm I)=0.\label{lindblad_condition}
\end{equation}
It means that the Markovian dynamics does not affect the maximally
mixed state $\mathrm I/2$. From Eq. (\ref{lindblad_condition}), we
arrive at the condition for the coefficient matrix
\begin{equation}
\gamma_{\alpha\beta}=\gamma_{\beta\alpha},\ \alpha,\beta=1,2,3.
\end{equation}
Notice that $\gamma=\gamma^{\dagger}$ should be hold because of the
Hermiticity of super-operator $\mathrm L$. Therefore, we can say
that the Markovian process described as Eq.
(\ref{lindblad_operator}) can create quantum discord for some input
semi-classical state at a finite time as long as at least one
coefficient $\gamma_{\alpha\beta}(\alpha,\beta\neq0)$ is not real.

Theorem 1 implies that mixedness does not contribute to quantum
discord for a two-qubit state. In the following, we will show that
mixedness can also contribute to quantum discord for states in
higher dimensions. Consider the following class of channels
\begin{equation}
\Lambda^{\mathrm U}(\cdot)\equiv\sum_ie_i^2u_i(\cdot)u_i^{\dagger},
\end{equation}
where $u_i$ are unitary operators and $\sum_ie_i^2=1$. Apparently,
$\{\Lambda^{\mathrm U}\}$ belongs to $\{\Lambda^{\mathrm M}\}$,
since
\begin{equation}
S(\Lambda^{\mathrm U}(\rho))\geq\sum_ie_i^2S(u_i\rho
u_i^{\dagger})=S(\rho).
\end{equation}
Now we will give an example to show that some $\Lambda^{\mathrm U}$
can create quantum deficit in classically correlated states.
Consider a channel $\Lambda^{\mathrm U}$ with Kraus operators
\begin{equation}
\mathrm E^{(0)}=e_0\mathrm I_3,\ \mathrm
E^{(1)}=e_1\left(\begin{array}{ccc}
\frac12&\frac12&\frac{1}{\sqrt2}\\
\frac12&\frac12&-\frac{1}{\sqrt2}\\
\frac{1}{\sqrt2}&-\frac{1}{\sqrt2}&0
\end{array}\right).
\end{equation}
A direct calculation can leads to
\begin{eqnarray}
&&[\Lambda^{\mathrm U}(|0\rangle\langle0|),\Lambda^{\mathrm
U}(|1\rangle\langle1|)]\nonumber\\
&=&e_1^2e_2^2\left(\begin{array}{ccc}
0&\frac12&-\frac{1}{2\sqrt2}\\
-\frac12&0&-\frac{1}{2\sqrt2}\\
\frac{1}{2\sqrt2}&\frac{1}{2\sqrt2}&0
\end{array}\right)\neq0.
\end{eqnarray}
Therefore, when
$\rho_{AB}=p_0\rho_A^0\otimes|0\rangle_B\langle0|+p_1\rho_A^1\otimes|1\rangle_B\langle1|$
is chosen as the initial state, the channel $\Lambda^{\mathrm U}$
can create quantum deficit between the qutrits $A$ and $B$. This
result shows that even a channel that can only cause mixing effect
can create the quantum discord.

We further prove that mixing channel can not increase the
teleportation fidelity of any two-qubit state. The average
teleportation fidelity $f$ is related to the maximum singlet
fraction (MSF) $F=\max_{\Phi}\langle\Phi|\rho|\Phi\rangle$ as
$f=(dF+1)/(d+1)$. After the action of mixing channel on $B$, the MSF
becomes
\begin{equation}
F'=\mathrm{Tr}(\rho\Xi)
\end{equation}
where $\Xi=\sum_i\mathrm I\otimes\mathrm
E^{(i)\dagger}|\Phi\rangle\langle\Phi|\mathrm I\otimes\mathrm
E^{(i)}$. Notice that for a mixing channel
$\Lambda(\cdot)=\sum_i\mathrm E^{(i)}(\cdot)\mathrm E^{(i)\dagger}$,
its conjecture $\Lambda^*(\cdot)=\sum_i\mathrm
E^{(i)\dagger}(\cdot)\mathrm E^{(i)}$ is also a mixing channel.
Therefore, $\Xi_A=\Xi_B=\mathrm I/2$, so $\Xi$ can be decomposed as
a mixture of maximally entangled pure states
$\Xi=\sum_ip_i|\Phi_i\rangle\langle\Phi_i|$. Then we have
$F'=\sum_ip_i\langle\Phi_i|\rho|\Phi_i\rangle\leq F$. Therefore,
average teleportation fidelity can never be increased by mixing
channel. This result suggests that quantum discord created by mixing
channel may not be a useful resource for quantum information tasks.

In summary, we have studied the condition for creation of quantum
correlations by local noises. For two-qubit half-classical states,
the necessary and sufficient condition for locally created quantum
discord is that the channel is neither a mixing channel nor a
completely decohering channel. For states in higher dimensions, we
give an example to show that even the mixing channel can create
quantum discord in some initially classical states. It means that
mixedness is important for quantum correlations in high-dimension
quantum states.

This work is supported NSFC under grants Nos. 10934010, 60978019,
the NKBRSFC under grants Nos. 2009CB930701, 2010CB922904,
2011CB921502, 2012CB821300, and NSFC-RGC under grants Nos.
11061160490 and 1386-N-HKU748/10.

Note added: in submitting our paper, we just notice a new online PRL
paper \cite{PhysRevLett.107.170502} which has similar conclusions
about the mixing channel and the completely decohering channel. Here
our methods are different from that paper.

\newpage %Just because of unusual number of tables stacked at end
\bibliography{apssamp}% Produces the bibliography via BibTeX.

\end{document}